\input{aipcheck}

\documentclass[
    ,final            
  ]
  {aipproc}

\layoutstyle{8x11double}

\begin{document}
\pdfoutput=1
\title{MANX:  A 6D 
Ionization-Cooling 
Experiment\thanks{To appear in {\sl Proc.\ NuFact07 Workshop}, Okayama, Japan (2007).}}


\author{Daniel M. Kaplan (for the MANX Collaboration)}{
  address={Illinois Institute of Technology, Chicago, IL 60616, USA}
}

\begin{abstract}
Six-dimensional ionization cooling of muons is essential for muon colliders and possibly beneficial for neutrino factories. An experiment to demonstrate six-dimensional ionization cooling using practical apparatus is presented. It exploits recent innovative ideas that may lead to six-dimensional muon-cooling channels with emittance reduction approaching that needed for high-luminosity muon colliders.
\end{abstract}

\maketitle

%%%%%%%%%%%%%%%%%%%%%%%%%%%%%%%%%%%%%%%%%%%%
%% MAINMATTER
%%%%%%%%%%%%%%%%%%%%%%%%%%%%%%%%%%%%%%%%%%%%

\section{Introduction}

Ionization cooling~\cite{cooling}, in which a beam is cooled by energy loss in an absorber medium, is a key technique for future muon accelerator facilities, e.g., a neutrino factory~\cite{factory} or muon collider~\cite{collider}. It is unique in its ability to cool an intense beam of muons before a substantial fraction of them have decayed. Ionization cooling is essentially a transverse effect but can be made to cool the longitudinal degrees of freedom as well via emittance exchange~\cite{Neuffer-yellow}. An experiment to demonstrate transverse ionization cooling (the Muon Ionization Cooling Experiment, MICE)~\cite{MICE} is in progress. We describe a possible six-dimensional (6D) cooling experiment: the Muon-collider And Neutrino-factory eXperiment, MANX~\cite{MANX}. 

\section{Six-Dimensional Muon Cooling}

Several approaches to six-dimensional muon cooling have been devised. The first design shown to work in simulation was the Balbekov ring cooler~\cite {Balbekov-ring}. Since then, several ring cooler designs have been studied, based on solenoid-focused ``RFOFO''  cells~\cite{RFOFO-ring} and quadrupole-~\cite{quad-ring} or dipole-edge-field-focused~\cite{edge-ring} cells. All can produce useful  levels of 6D cooling, but injection and extraction are problematic. This problem is eliminated (at the expense of greater hardware cost) by extending an RFOFO ring into the third dimension, giving a helical, ``Guggenheim''  cooling channel~\cite{Guggenheim}.\footnote{The allusion is to the Guggenheim Museum in New York, rather than, say, that in Bilbao.} This can also alleviate problematic RF loading and absorber heating, and it allows the focusing strength at each step along the device to be tailored to the emittance at that point, enhancing the cooling efficacy. In all of these designs, bending magnets introduce the dispersion needed for longitudinal--transverse emittance exchange.

\subsection{Helical Cooling Channel}

\begin{figure}
  \includegraphics[width=.7\textwidth]{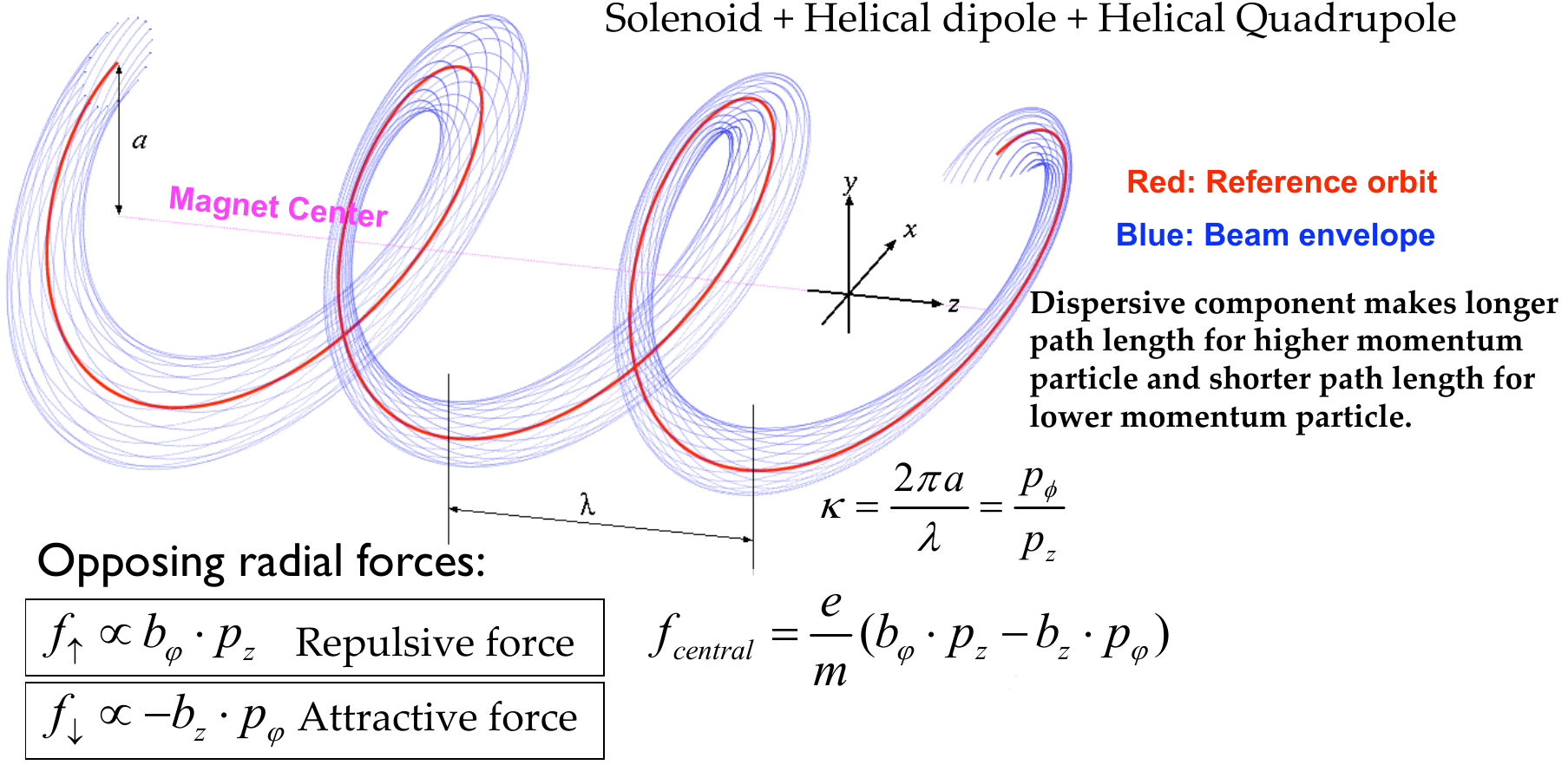}
  \includegraphics[width=.23\textwidth, bb=20 40 380 560,clip]{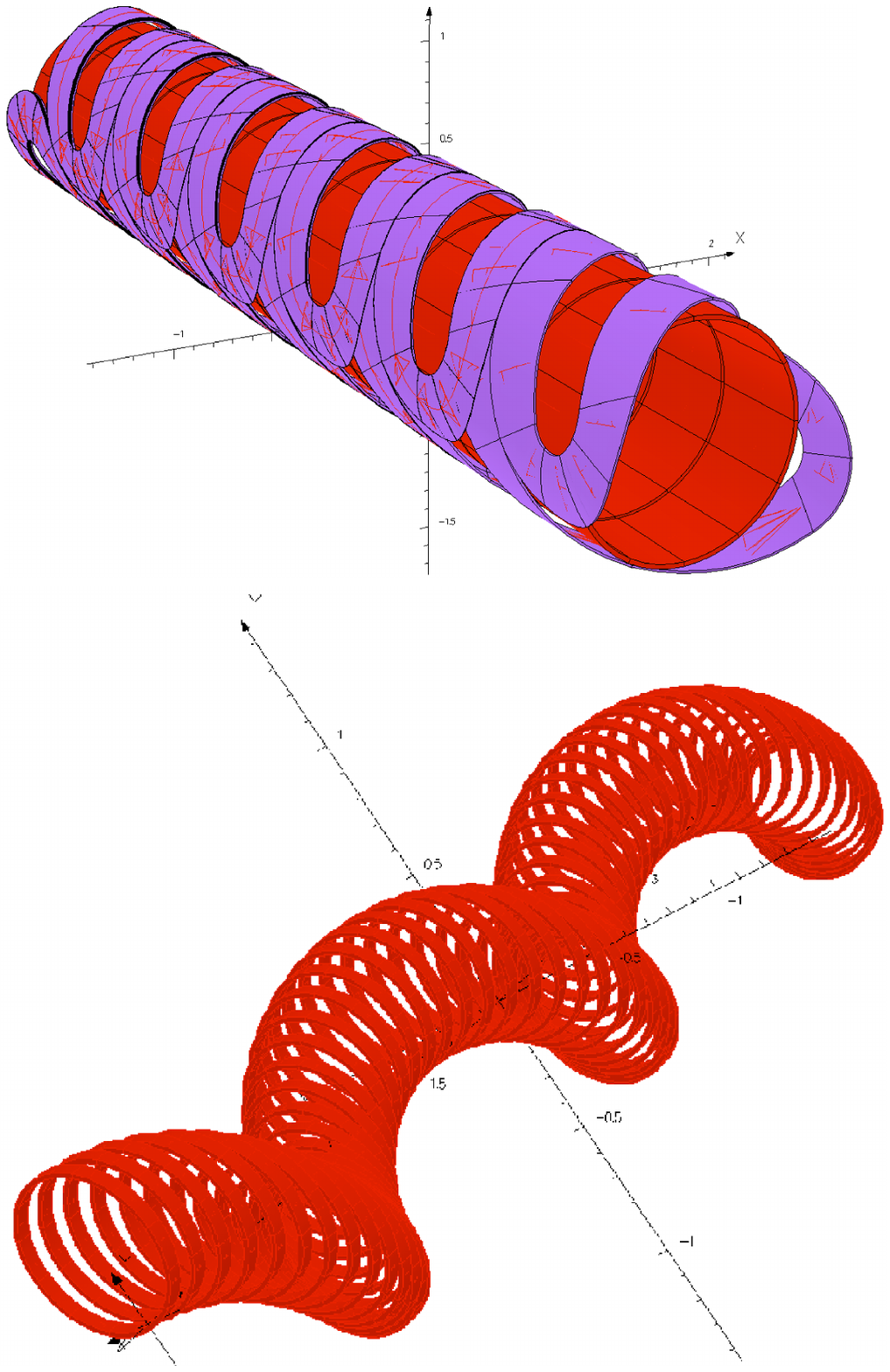}
  \caption{(left) Helical-channel principle; (top-right) conventional ``Siberian snake'' solution with individual windings providing the required solenoidal, helical dipole, and helical quadrupole fields; (bottom-right) Helical Solenoid implementation with the same acceptance and the three required fields produced using
simple offset coils only half the diameter of the conventional magnet.}
  \label{fig-helix}
\end{figure}

A more recent development is the Helical Cooling Channel (HCC)~\cite{Derbenev-Johnson},  employing  a helical dipole field superimposed on a solenoid field. The helical dipole, 
known from ``Siberian Snake'' magnets used to control spin resonances in synchrotrons, provides the dispersion needed for emittance exchange.  The solenoid field 
provides focusing, and helical quadrupole magnets are added for beam stability and larger acceptance.  Figure~\ref{fig-helix} illustrates the beam motion, as well as two possible magnet configurations: a conventional one with three separate windings generating the required field components, and the recent ``Helical Solenoid'' invention~\cite{Kashikhin-etal}, which achieves the same field components and acceptance using simple circular coils of half the radius, 
about one-quarter the stored energy, and smaller fields at the conductors.  The equilibrium  beam orbit  follows the centers of the coils. (The theory of the HCC, based on a Hamiltonian formalism that starts with the opposing radial forces shown in Fig. 1, is derived in~\cite{Derbenev-Johnson}.)
 
\subsection{Continuous Absorber}

\begin{figure}
  \includegraphics[width=.479\textwidth, bb=10 0 535 269,clip]{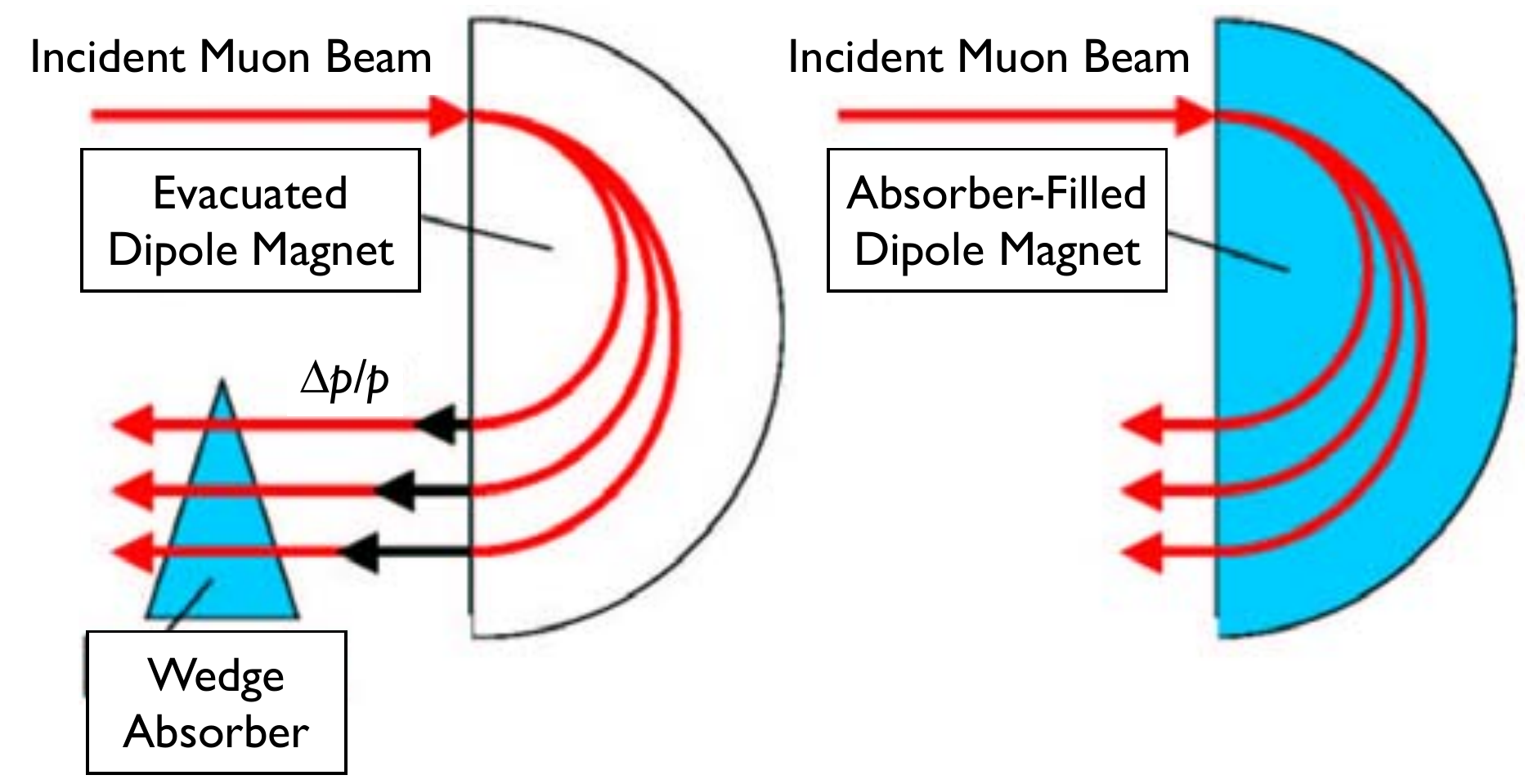}
    \caption{(left) Emittance exchange via dispersion and wedge absorber; (right) emittance exchange via continuous absorber.}
  \label{fig-wedge}
\end{figure}

Six-dimensional muon coolers were first formulated with emittance exchange via wedge absorbers located at dispersive points in the lattice. The same effect may be achieved more simply by use of a continuous absorber~\cite{Derbenev-Johnson,Recent} (Fig.~\ref{fig-wedge}). This approach may be synergistic with the idea of maximizing the operating gradient of copper RF cavities in high magnetic fields by filling them with pressurized hydrogen~\cite{gas-filled}; the absorber needed for ionization cooling can thus be combined with the muon re-acceleration, giving a shorter and more adiabatic channel. Another possibility is a ``separated-function'' cooling channel in which pressurized-gas- or liquid-filled HCC segments are separated by linear-accelerator sections; in such an arrangement, the fields of each HCC segment can be graded~\cite{Recent}, so as to maintain constant focusing strength as  the beam momentum is reduced by energy loss 
in the absorber medium. Such an arrangement may be advantageous in that the acceleration could then be done using superconducting RF cavities, reducing instantaneous-power requirements.

\subsection{HCC Example}

\begin{figure}
  \includegraphics[width=.475\textwidth, bb=10 5 542 635,clip]{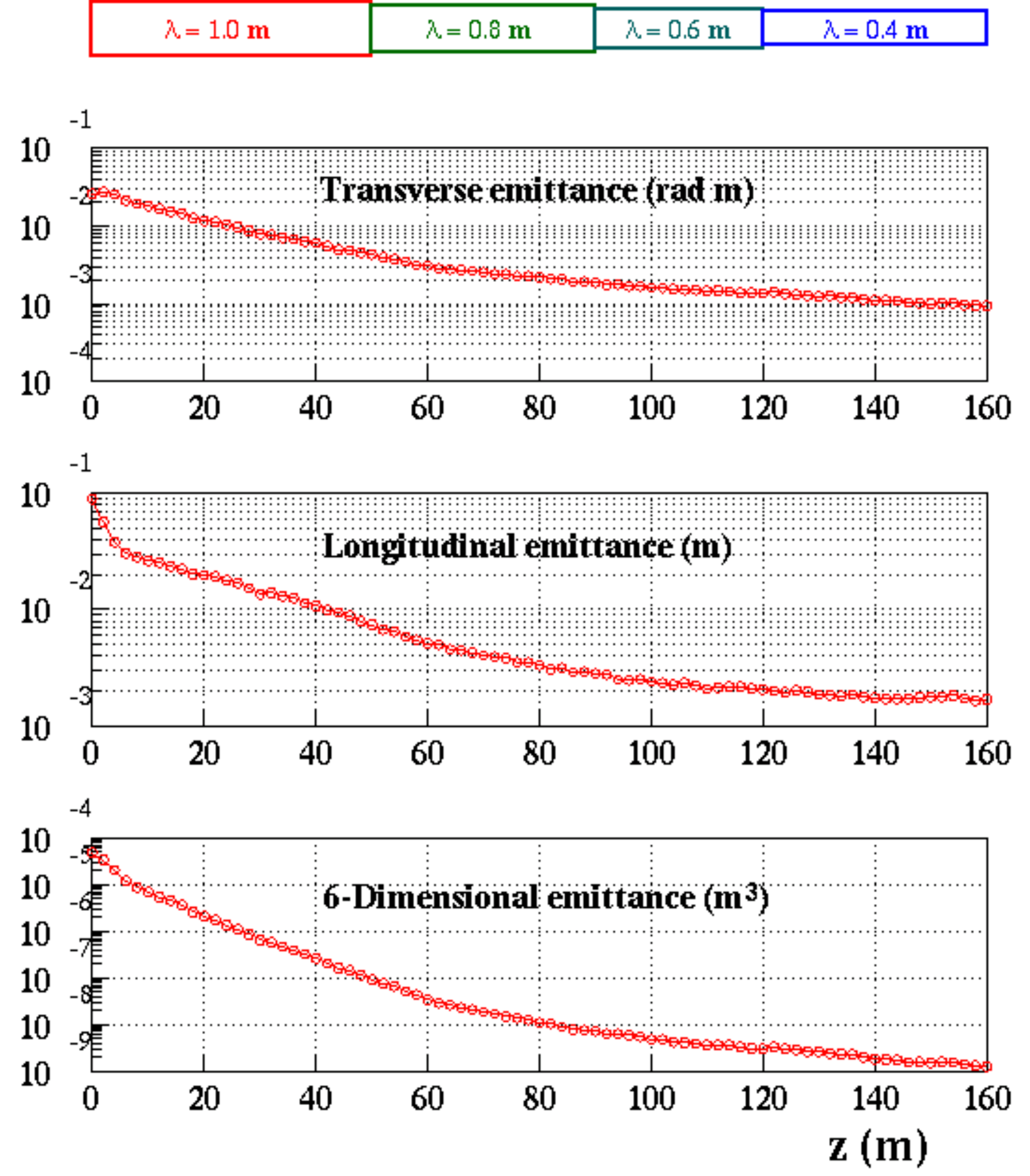}
    \caption{Simulation of emittance reduction in a 4-segment, 160\,m HCC filled with high-pressure hydrogen gas.}
  \label{fig-example}
\end{figure}

Figure~\ref{fig-example} shows the results of a G4beamline~\cite{G4BL} simulation of a 160\,m, 4-section HCC carried out by K. Yonehara~\cite{Recent}. The 6D emittance reduction factor of $5\times10^4$ is a big step towards the $\sim$10$^6$ required for a high-luminosity muon collider. Cooling approaches capable of providing the additional factor of 10--100 needed  are under development~\cite{Recent}.

\section{MANX}

These innovative muon-cooling approaches will require experimental demonstration before a facility employing them can be approved for construction. Since such demonstrations are potentially expensive (typically comparable in cost to medium-scale HEP experiments), which aspects to demonstrate, and how best to do so, must be considered with care.

\begin{figure}
\centerline{\includegraphics[width=0.48\textwidth, bb=20 40 755 560, clip]{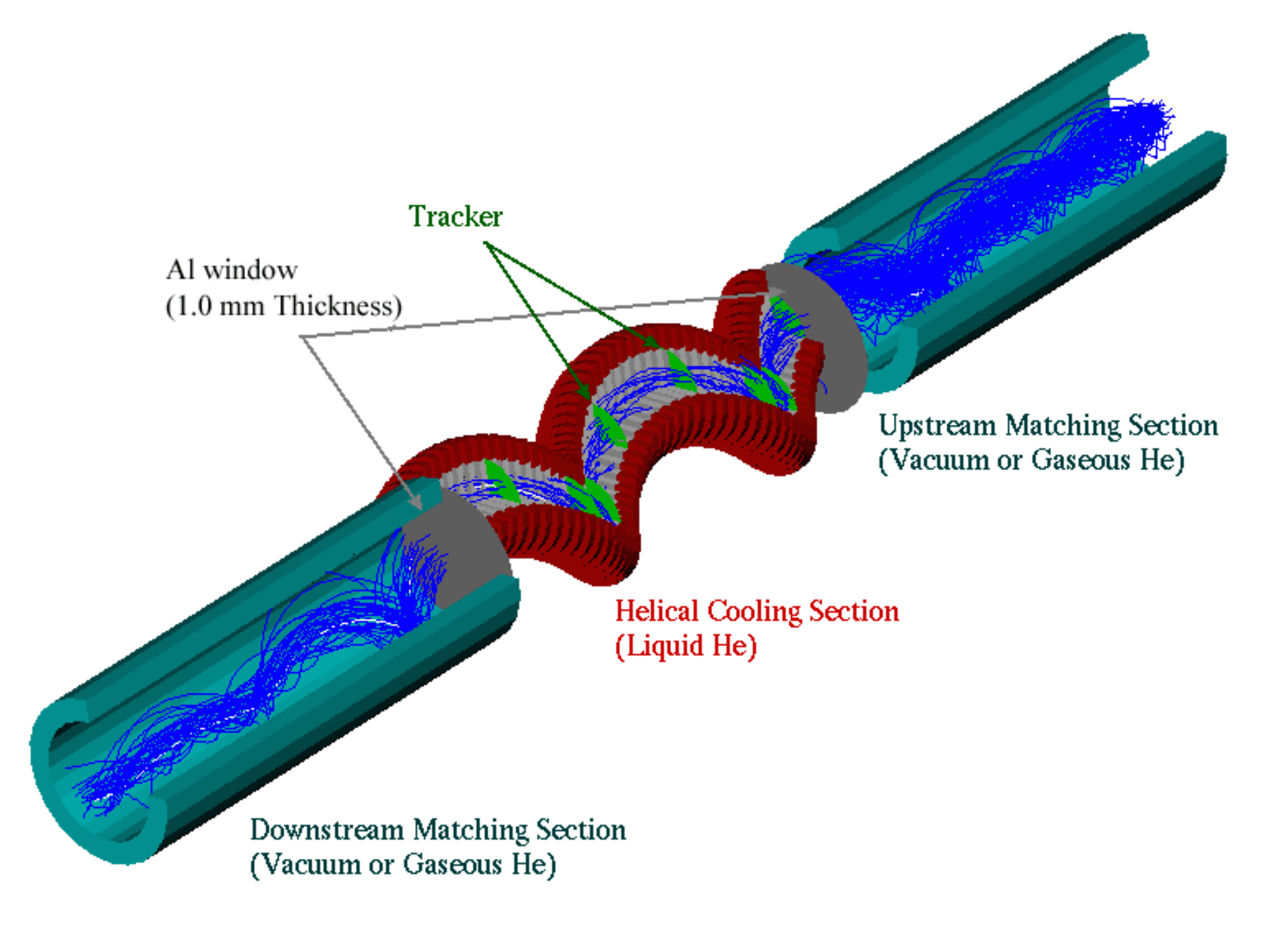}}
\caption{Simulation of possible MANX HCC section between matching sections. The solenoid and rotating-dipole fields gradually turn on (off)  in the upstream (downstream) matching section. The overall length in this example is 12\,m.}\label{fig:MANX}
\end{figure}

A proposal for a 6D HCC demonstration experiment is under development by a collaboration among Muons, Inc.~\cite{muonsinc}, Fermilab, and university groups~\cite{MANX}. The approach taken is to design a separated-function, graded-HCC segment of modest length which nevertheless delivers an impressive ($\approx 3$--5) 6D cooling factor. Such a device might be suitable for use as a precooler to a combined-function HCC incorporating pressurized RF cavities, or as a first segment in a separated-function HCC. It may also be capable of increasing substantially the rate of muons stopping in a thin target, e.g., in a muon-to-electron-conversion experiment~\cite{mu2e}.

By eliminating the RF cavities, the cost is substantially reduced and the attention is focused on the dynamics and engineering issues of the HCC magnet itself. While this is not the only approach that might be taken in such a demonstration experiment, it may be a sensible one in that it ``factorizes'' the engineering challenges:  with hydrogen-absorber operation in close proximity to RF cavities and high-field solenoids already being tackled by MICE, arguably this need not be demonstrated again before  a full muon accelerator facility is engineered.
 
The MANX apparatus will include muon-measure\-ment sections and  (Fig.~\ref{fig:MANX}) matching sections into and out of the cooling section; it may also be possible to operate thin tracking detectors within the HCC section as indicated in Fig.~\ref{fig:MANX}.
 
Various venues for MANX are being explored. The MICE muon beamline and detectors might be re-usable for MANX; options involving a new muon beam at Fermilab are also under consideration. It is hoped to carry out the experiment within the next $\stackrel{<}{_\sim}$ 5 years.

\begin{theacknowledgments}
I thank my collaborators at Muons, Inc., who devised many of the innovations discussed here. This work is supported by the US Dept. of Energy. The G4beamline program is in ongoing development by IIT and Muons, Inc. under STTR grant DE-FG02-06ER86281.
\end{theacknowledgments}

\bibliographystyle{aipproc}   
\bibliography{sample}

\IfFileExists{\jobname.bbl}{}
 {\typeout{}
  \typeout{******************************************}
  \typeout{** Please run ''bibtex \jobname'' to optain}
  \typeout{** the bibliography and then re-run LaTeX}
  \typeout{** twice to fix the references!}
  \typeout{******************************************}
  \typeout{}
 }

\end{document}